\documentclass[10pt,english]{article}
\usepackage{times}
\usepackage[T1]{fontenc}
\usepackage[latin1]{inputenc}
\usepackage{amsmath}
\usepackage{amssymb}

\makeatletter

\providecommand{\tabularnewline}{\\}

\usepackage{slashed}

\usepackage{babel}
\makeatother
\begin{document}

\title{\textbf{CANONICAL SEESAW MECHANISM IN ELECTRO-WEAK $SU(4)_{L}\otimes U(1)_{Y}$
MODELS}}

\author{ADRIAN PALCU}

\date{\emph{Faculty of Exact Sciences - {}``Aurel Vlaicu'' University
Arad, Str. Elena Dr\u{a}goi 2, Arad - 310330, Romania}}

\maketitle
\begin{abstract}
In this paper we prove that the canonical seesaw mechanism can naturally
be implemented in a particular class of electro-weak \textbf{$SU(4)_{L}\otimes U(1)_{Y}$}
gauge models. The resulting neutrino mass spectrum is determined by
just tuning a unique free parameter $a$ within the algebraical method
of solving gauge models with high symmetries. All the Standard Model
phenomenology is preserved, being unaffected by the new physics occuring
at a high breaking scale $m\sim10^{11}$GeV.

PACS numbers: 14.60.St; 14.60.Pq; 12.60.Fr.

Key words: 3-4-1 gauge models, see-saw mechanism, neutrino masses. 
\end{abstract}

\section{Introduction}

One of the main reasons driving the search for various extensions
of the Standard Model (SM) \cite{key-1} - \cite{key-3} - which has
been established as a gauge theory based on the local group $SU(3)_{C}\otimes SU(2)_{L}\otimes U(1)_{Y}$
undergoing in its electro-weak sector a spontaneous symmetry breakdown
(SSB) up to the electromagnetic universal $U(1)_{em}$ - stems in
the recently unfolded neutrino phenomenology \cite{key-4}. At the
SM level, it is well known that left-handed neutrinos are massless
in all orders of perturbations and there is no need at all for right-handed
neutrinos. Therefore, no mixing occurs in the lepton sector, in contrast
with the quark sector. Observational collaborations such as SuperKamiokande
\cite{key-5,key-6}, K2K \cite{key-7}, SNO \cite{key-8}, KamLAND
\cite{key-9}, LSND \cite{key-10} and others have definitely proved
within the last decade that neutrinos oscillate and, consequently,
they must carry non-zero masses. Although the mass spectrum in the
neutrino sector exhibits certain features such as the mass splitting
ratio $r_{\Delta}=\Delta m_{\odot}^{2}/\Delta m_{atm}^{2}\sim0.03$
and particular mixing angles ($\theta_{\odot}\simeq34^{\circ}$ and
$\theta_{\odot}\simeq45^{\circ}$, along with $\theta_{13}\simeq0$
in the mixing matrix), the absolute mass hierarchy has not been determined
yet. What we only know at present is that it lies most likely in the
$eV$ region.

However, the theoretical devices designed to face such a reality (see
Refs. \cite{key-11} - \cite{key-13} for excellent reviews on neutrino
mass issue) mainly included two distinct purposals - (i) radiative
mechanisms (initially proposed by Zee \cite{key-14}) and (ii) various
types of see-saw \cite{key-15} - \cite{key-18} - in order to obtain
viable predictions for the massive neutrino sector. Notwithstanding,
these approaches seem more efficient in some extentions of the SM,
since models such as $SU(3)_{C}\otimes SU(3)_{L}\otimes U(1)_{Y}$
(3-3-1) introduced and developed by Frampton, Pisano and Pleitez \cite{key-19},
Frampton \cite{key-20}, Long \cite{key-21} and others \cite{key-22}
- \cite{key-33} and $SU(3)_{C}\otimes SU(4)_{L}\otimes U(1)_{Y}$
(3-4-1) \cite{key-34} - \cite{key-48} have emerged in the literature.
Neutrino masses generated through radiative patterns in 3-3-1 models
can be found in Refs. \cite{key-49} - \cite{key-56}, while the way
see-saw mechanisms work in those models is exploited in Refs. \cite{key-57}
- \cite{key-61}.

Here we are concerned with the well-known see-saw mechanism \cite{key-15}
- \cite{key-17}) worked out in the particular electro-weak $SU(4)_{L}\otimes U(1)_{Y}$
model without exotic electric charges \cite{key-44} - \cite{key-47}.
As a matter of fact, this efficient mathematical procedure calls for
both left-handed and right-handed neutrinos. Since they can naturally
be embedded in lepton multiplets of the 3-4-1 model, there is no need
for supplemental ingredients (like a new small parameter as in Ref.
\cite{key-59}). 

The paper is organized as follows: Sec. 2 reviews the main features
of the $SU(3)_{C}\otimes SU(4)_{L}\otimes U(1)_{Y}$ gauge model without
exotic electric charges treated within the framework of the general
algebraical method proposed by Cot\u{a}escu \cite{key-62}. Sec.
3 deals with the fermion mass issue with a special emphasize on the
neutrino Yukawa sector in the 3-4-1 model of interest. Our conclusions
are sketched in the last section (Sec. 4) where also some numerical
estimates are given.

\section{$SU(4)_{L}\otimes U(1)_{Y}$ model without exotic charges}

All the details of the of solving gauge models with high symmetries
undergoing in their electro-weak a SSB can be found in the paper of
Cot\u{a}escu \cite{key-62}. Note that it was already succesfully
applied by the author in the case of 3-3-1 gauge models in a series
of papers \cite{key-63} - \cite{key-67} which accomodated in a natural
way the neutrino phenomenology. In each of these cases, the method
itself led to viable predictions regarding the boson mass spectrum
and currents (both charged and neutral), recovering all the SM phenomenology.
This goal was simply achieved by introducing a particular metric in
the scalar sector that finally offered a framework with a single free
parameter to be tuned. The same procedure was exploited in recent
papers dealing with the 3-4-1 models \cite{key-43,key-47} with a
remarkable succes with regard to the boson mass spectrum and coupling
coefficients for the charged and neutral currents (at least in the
no-exotic-electric-charges class of 3-4-1 models \cite{key-47}).
In this latter case we go further by introducing a special arrangment
in the Higgs sector leading to a natural see-saw context in the neutrino
sector. 

In this secton we briefly present the particle content of the 3-4-1
model of interest here, namely Model A in Ref. \cite{key-40}. For
the $SU(4)_{L}$ group the 3 diagonal generators are defined as: $D_{1}=T_{3}=\frac{1}{2}Diag(1,-1,0,0)$,
$D_{2}=T_{8}=\frac{1}{2\sqrt{3}}Diag(1,1,-2,0)$, and $D_{3}=T_{15}=\frac{1}{2\sqrt{6}}Diag(1,1,1,-3)$
respectively. The irreducible representations (irreps) with respect
to the gauge group of the theory are denoted by $(\mathbf{n}_{color},\mathbf{n}_{\rho},y_{ch}^{\rho})$
while the versor assignment needed in the general method \cite{key-62}
stands as $\nu_{1}=0$, $\nu_{2}=0$, $\nu_{3}=-1$. The parameter
matrix \cite{key-62} in the scalar sector is taken as $\eta^{2}=(1-\eta_{0}^{2})Diag\left(1-c,c-a,\frac{1}{2}a+b,\frac{1}{2}a-b\right)$
in order to fulfil the condition $Tr(\eta^{2})=1-\eta_{0}^{2}$ in
the general method. At the same time, one assumes the condition $e=g\sin\theta_{W}$
established in the SM and the relation between $\theta_{W}$ and $\theta$
(introduced by the method itself in order to separate the electromagnetic
field in a general Weinberg transformation - see Sec. 5 in Ref. \cite{key-62})
yields $\sin\theta=\sqrt{\frac{3}{2}}\sin\theta_{W}$. Under these
circumstances, the coupling matching was inferred \cite{key-40} on
algebraical reasons: $\frac{g^{\prime}}{g}=\frac{\sin\theta_{W}}{\sqrt{1-\frac{3}{2}\sin^{2}\theta_{W}}}$
(where, obviously, $g$ is the $SU(4)_{L}$ coupling and $g^{\prime}$
is the $U(1)_{em}$ coupling).

\subsection{Fermion content}

The fermion sector of the Model A \cite{key-40} consists of the following
representations:

\textbf{Lepton families}\begin{equation}
\begin{array}{ccccc}
f_{\alpha L}=\left(\begin{array}{c}
N_{\alpha}^{\prime}\\
N_{\alpha}\\
\nu_{\alpha}\\
e_{\alpha}\end{array}\right)_{L}\sim(\mathbf{1,4^{*}},-1/4) &  &  &  & \left(e_{\alpha L}\right)^{c}\sim(\mathbf{1,1},1)\end{array}\label{Eq.1}\end{equation}

\textbf{Quark families}\begin{equation}
\begin{array}{ccc}
Q_{iL}=\left(\begin{array}{c}
D_{i}^{\prime}\\
D_{i}\\
-d_{i}\\
u_{i}\end{array}\right)_{L}\sim(\mathbf{3,4},-1/12) &  & Q_{3L}=\left(\begin{array}{c}
U^{\prime}\\
U\\
u_{3}\\
d_{3}\end{array}\right)_{L}\sim(\mathbf{3},\mathbf{4^{*}},5/12)\end{array}\label{Eq.2}\end{equation}
\begin{equation}
\begin{array}{c}
(d_{3L})^{c},(d_{iL})^{c},(D_{iL})^{c},(D_{iL}^{\prime})^{c}\sim(\mathbf{3},\mathbf{1},+1/3)\end{array}\label{Eq.3}\end{equation}

\begin{equation}
(u_{3L})^{c},(u_{iL})^{c},(U_{L})^{c},(U_{L}^{\prime})^{c}\sim(\mathbf{3},\mathbf{1},-2/3)\label{Eq.4}\end{equation}
with $\alpha=1,2,3$ and $i=1,2$. 

With this assignment the fermion families cancel the axial anomalies
by just an interplay between them, although each family remains anomalous
by itself.

\subsection{Boson sector}

The boson sector is determined by the standard generators $T_{a}$
of the $su(4)$ algebra. In this basis, the gauge fields are $A_{\mu}^{0}$
and $A_{\mu}\in su(4)$, that is \begin{equation}
A_{\mu}=\frac{1}{2}\left(\begin{array}{ccccccc}
D_{\mu}^{1} &  & \sqrt{2}Y_{\mu} &  & \sqrt{2}X_{\mu}^{\prime} &  & \sqrt{2}X_{\mu}^{\prime}\\
\\\sqrt{2}Y_{\mu}^{*} &  & D_{\mu}^{2} &  & \sqrt{2}K_{\mu} &  & \sqrt{2}K_{\mu}^{\prime}\\
\\\sqrt{2}X_{\mu}^{*}{} &  & \sqrt{2}K_{\mu}^{*} &  & D_{\mu}^{3} &  & \sqrt{2}W_{\mu}\\
\\\sqrt{2}X_{\mu}^{\prime*} &  & \sqrt{2}K_{\mu}^{\prime*} &  & \sqrt{2}W_{\mu}^{*} &  & D_{\mu}^{4}\end{array}\right),\label{Eq.5}\end{equation}
with $D_{\mu}^{1}=A_{\mu}^{3}+A_{\mu}^{8}/\sqrt{3}+A_{\mu}^{15}/\sqrt{6}$,
$D_{\mu}^{2}=-A_{\mu}^{3}+A_{\mu}^{8}/\sqrt{3}+A_{\mu}^{15}/\sqrt{6}$,
$D_{\mu}^{3}=-2A_{\mu}^{8}/\sqrt{3}+A_{\mu}^{15}/\sqrt{6}$, $D_{\mu}^{4}=-3A_{\mu}^{15}/\sqrt{6}$
as diagonal bosons. Apart from the charged Weinberg bosons ($W^{\pm}$),
there are two new charged bosons, $K^{0}$, $K^{\prime\pm}$, while
$X^{0}$, $X^{\prime\pm}$ and $Y^{0}$ are new neutral bosons, but
distinct from the diagonl ones ($Z$, $Z^{\prime}$, $Z^{\prime\prime}$
plus the massless $A_{em}$).

\subsection{Minimal Higgs mechanism}

The general method assumes also a particular minimal Higgs mechanism
(mHm) based on a special parametrization in the scalar sector, such
that the $n$ Higgs multiplets $\phi^{(1)}$, $\phi^{(2)}$, ... $\phi^{(n)}$
satisfy the orthogonality condition $\phi^{(i)+}\phi^{(j)}=\phi^{2}\delta_{ij}$
in order to eliminate the unwanted Goldstone bosons that could survive
the SSB. $\phi$ is a gauge-invariant real scalar field while the
Higgs multiplets $\phi^{(i)}$ transform according to the irreps $(\mathbf{1},\mathbf{n},y^{(i)})$
whose characters $y^{(i)}$ are arbitrary numbers that can be organized
into the diagonal matrix $Y=Diag\left(y^{(1)},y^{(2)},\cdots,y^{(n)}\right)$.
The Higgs sector needs, in our approach, a parameter matrix \begin{equation}
\eta=Diag\left(\eta{}^{(1)},\eta{}^{(2)},...,\eta{}^{(n)}\right)\label{Eq.6}\end{equation}
 with the property $Tr(\eta^{2})=1-\eta_{0}^{2}$. It will play the
role of the metric in the kinetic part of the Higgs Lagrangian density
(Ld) which reads \begin{equation}
\mathcal{L}_{H}=\frac{1}{2}\eta_{0}^{2}\partial_{\mu}\phi\partial^{\mu}\phi+\frac{1}{2}\sum_{i=1}^{n}\left(\eta{}^{(i)}\right)^{2}\left(D_{\mu}\phi^{(i)}\right)^{+}\left(D^{\mu}\phi^{(i)}\right)-V(\phi)\label{Eq.7}\end{equation}
 where $D_{\mu}\phi^{(i)}=\partial_{\mu}\phi^{(i)}-ig(A_{\mu}+y^{(i)}A_{\mu}^{0})\phi^{(i)}$
are the covariant derivatives of the model and $V(\phi)$ is the scalar
potential generating the SSB of the gauge symmetry \cite{key-62}.
This is assumed to have an absolute minimum for $\phi=\langle\phi\rangle\not=0$
that is, $\phi=\langle\phi\rangle+\sigma$ where $\sigma$ is the
unique surviving physical Higgs field. Therefore, one can always define
the unitary gauge where the Higgs multiplets, $\hat{\phi}^{(i)}$
have the components $\hat{\phi}_{k}^{(i)}=\delta_{ik}\phi=\delta_{ik}(\langle\phi\rangle+\sigma)$.

The masses of both the neutral and charged bosons depend on the choice
of the matrix $\eta$ whose components are free parameters. Here it
is convenient to assume the following matrix \begin{equation}
\eta^{2}=(1-\eta_{0}^{2})Diag\left(1-c,c-a,\frac{1}{2}a+b,\frac{1}{2}a-b\right),\label{Eq.8}\end{equation}
 where, for the moment, $a$,$b$ and $c$ are arbitrary non-vanishing
real parameters. Obviously, $\eta_{0},c\in[0,1)$, $a\in(0,c)$ and
$b\in(-a,+a)$. 

With this assignment - for all the details the reader is referred
to Ref.\cite{key-47} - after some algebra exploiting the mass relation
from SM $m^{2}(Z)=m^{2}(W)/\cos^{2}\theta_{W}$ (equivalent with $Det\left|M^{2}-\frac{m^{2}a}{\cos^{2}\theta_{W}}\right|=0$)
and enforcing some physical arguments in the above presented 3-4-1
model regarding the decoupling of the heaviest $Z^{\prime\prime}$
as the symmetry is broken to $SU(3)$ (equivalent with $c=(1+a)/2$),
one obtains a one-parameter mass scale (by working out the relation$b=\frac{1}{2}a\tan^{2}\theta_{W}$)
(see Sec. 2.4). 

It is worth noting that the parameter matrix now becomes

\begin{equation}
\eta^{2}=(1-\eta_{0}^{2})Diag\left(\frac{1-a}{2},\frac{1-a}{2},\frac{a}{2}(1+\tan^{2}\theta_{W}),\frac{a}{2}(1-\tan^{2}\theta_{W}),\right),\label{Eq.9}\end{equation}
while the 4 scalar 4-plets of the Higgs sector are represented by
$\phi^{(1)},\phi^{(2)},\phi^{(3)}\sim(\mathbf{1,4},1/4)$ and $\phi^{(4)}\sim(\mathbf{1,4},-3/4)$.
They can be re-defined as $\phi^{(i)}\rightarrow\eta^{(i)}\phi^{(i)}$without
altering the physical content (as one can see in Sec. 3).

\subsection{Boson mass spectrum}

With the following notation $m^{2}=g^{2}\left\langle \phi\right\rangle ^{2}(1-\eta_{0}^{2})/4$
the masses of the physical bosons stand

\begin{eqnarray}
{m}^{2}(W) & = & m^{2}a,\label{Eq10}\\
{m}^{2}(X) & = & m^{2}a\left(\frac{1+\tan^{2}\theta_{W}}{2}\right),\label{Eq.11}\\
{m}^{2}(X') & = & m^{2}a\left(\frac{1-\tan^{2}\theta_{W}}{2}\right),\label{Eq.12}\\
{m}^{2}(K) & ={} & m^{2}a\left(\frac{1+\tan^{2}\theta_{W}}{2}\right),\label{Eq.13}\\
{m}^{2}(K') & = & m^{2}a\left(\frac{1-\tan^{2}\theta_{W}}{2}\right),\label{Eq.14}\\
{m}^{2}(Y) & = & m^{2}(1-a),\label{Eq.15}\\
{m}^{2}(Z) & = & m^{2}a/\cos^{2}\theta_{W},\label{Eq.16}\\
{m}^{2}(Z^{\prime}) & = & {m}^{2}\frac{\cos^{4}\theta_{W}-a\sin^{4}\theta_{W}}{\cos^{2}\theta_{W}\left(2-3\sin^{2}\theta_{W}\right)},\label{Eq.17}\\
{m}^{2}(Z^{\prime\prime}) & = & m^{2}(1-a).\label{Eq.18}\end{eqnarray}

One can observe that the above mass scale is just a matter of tuning
the parameter $a$ in accordance with the possible values for $\left\langle \phi\right\rangle $.

\subsection{Neutral charges}

Now one can compute all the charges for the fermion representations
in model A with respect to the neutral bosons ($Z$, $Z^{\prime}$,
$Z^{\prime\prime}$), since the general Weinberg transformation (gWt)
is determined by the matrix

\begin{equation}
\omega=\left(\begin{array}{ccccc}
1 &  & 0 &  & 0\\
\\0 &  & \frac{1}{\sqrt{3}\sqrt{1-\sin^{2}\theta_{W}}} &  & \frac{\sqrt{2-3\sin^{2}\theta_{W}}}{\sqrt{3}\sqrt{1-\sin^{2}\theta_{W}}}\\
\\0 &  & -\frac{\sqrt{2-3\sin^{2}\theta_{W}}}{\sqrt{3}\sqrt{1-\sin^{2}\theta_{W}}} &  & \frac{1}{\sqrt{3}\sqrt{1-\sin^{2}\theta_{W}}}\end{array}\right).\label{Eq.19}\end{equation}

They will be expressed (assuming the above versor assignment $\nu_{1}=0$,
$\nu_{2}=0$, $\nu_{3}=-1$) by:

\begin{equation}
Q^{\rho}(Z^{\hat{i}})=g\left[D_{1}^{\rho}\omega_{\cdot\;\hat{i}}^{1\;\cdot}+D_{2}^{\rho}\omega_{\cdot\;\hat{i}}^{2\;\cdot}+\left(D_{3}^{\rho}\cos\theta+y_{ch}^{\rho}\frac{g^{\prime}}{g}\sin\theta\right)\omega_{\cdot\;\hat{i}}^{3\;\cdot}\right],\label{Eq.20}\end{equation}
where the conditions $\frac{g^{\prime}}{g}=\frac{\sin\theta_{W}}{\sqrt{1-\frac{3}{2}\sin^{2}\theta_{W}}}$
and $\sin\theta=\sqrt{\frac{3}{2}}\sin\theta_{W}$ have to be inserted.
The couplings are listed in the following Table. 

\begin{table}

\caption{Coupling coefficients of the neutral currents in 3-4-1 model }

\begin{tabular}{cccccc}
\hline 
Particle\textbackslash{}Coupling($e/\sin2\theta_{W}$)&
$Z\rightarrow\bar{f}f$&
&
$Z^{\prime}\rightarrow\bar{f}f$&
&
$Z^{\prime\prime}\rightarrow\bar{f}f$\tabularnewline
\hline
\hline 
&
&
&
&
&
\tabularnewline
&
&
&
&
&
\tabularnewline
$\nu_{eL},\nu_{\mu L},\nu_{\tau L}$&
$1$&
&
$\frac{1-3\sin^{2}\theta_{W}}{2\sqrt{2-3\sin^{2}\theta_{W}}}$&
&
$0$\tabularnewline
&
&
&
&
&
\tabularnewline
$e_{L},\mu_{L},\tau_{L}$&
$2\sin^{2}\theta_{W}-1$&
&
$\frac{1-3\sin^{2}\theta_{W}}{2\sqrt{2-3\sin^{2}\theta_{W}}}$&
&
$0$\tabularnewline
&
&
&
&
&
\tabularnewline
$N_{eL},N_{\mu L},N_{\tau L}$&
$0$&
&
$-\frac{3\cos^{2}\theta_{W}}{2\sqrt{2-3\sin^{2}\theta_{W}}}$&
&
$\cos\theta_{W}$\tabularnewline
&
&
&
&
&
\tabularnewline
$N_{eL}^{\prime},N_{\mu L}^{\prime},N_{\tau L}^{\prime}$&
$0$&
&
$-\frac{3\cos^{2}\theta_{W}}{2\sqrt{2-3\sin^{2}\theta_{W}}}$&
&
$-\cos\theta_{W}$\tabularnewline
&
&
&
&
&
\tabularnewline
$e_{R},\mu_{R},\tau_{R}$&
$2\sin^{2}\theta_{W}$&
&
$-\frac{2\sin^{2}\theta_{W}}{\sqrt{2-3\sin^{2}\theta_{W}}}$&
&
$0$\tabularnewline
&
&
&
&
&
\tabularnewline
$u_{L},c_{L}$&
$1-\frac{4}{3}\sin^{2}\theta_{W}$&
&
$\frac{2-9\cos^{2}\theta_{W}}{2\sqrt{2-3\sin^{2}\theta_{W}}}$&
&
$0$\tabularnewline
&
&
&
&
&
\tabularnewline
$d_{L},s_{L}$&
$-1+\frac{2}{3}\sin^{2}\theta_{W}$&
&
$\frac{2-9\cos^{2}\theta_{W}}{2\sqrt{2-3\sin^{2}\theta_{W}}}$&
&
0\tabularnewline
&
&
&
&
&
\tabularnewline
$t_{L}$&
$1-\frac{4}{3}\sin^{2}\theta_{W}$&
&
$\frac{2+9\cos^{2}\theta_{W}}{6\sqrt{2-3\sin^{2}\theta_{W}}}$&
&
0\tabularnewline
&
&
&
&
&
\tabularnewline
$b_{L}$&
$-1+\frac{2}{3}\sin^{2}\theta_{W}$&
&
$\frac{2+9\cos^{2}\theta_{W}}{6\sqrt{2-3\sin^{2}\theta_{W}}}$&
&
0\tabularnewline
&
&
&
&
&
\tabularnewline
$u_{R},c_{R},t_{R},U_{1R},U_{iR}^{\prime}$&
$-\frac{4}{3}\sin^{2}\theta_{W}$&
&
$\frac{4\sin^{2}\theta_{W}}{3\sqrt{2-3\sin^{2}\theta_{W}}}$&
&
$0$\tabularnewline
&
&
&
&
&
\tabularnewline
$d_{R},s_{R},b_{R},D_{iR},D_{iR}^{\prime}$&
$+\frac{2}{3}\sin^{2}\theta_{W}$&
&
$-\frac{2\sin^{2}\theta_{W}}{3\sqrt{2-3\sin^{2}\theta_{W}}}$&
&
$0$\tabularnewline
&
&
&
&
&
\tabularnewline
$D_{1L},D_{2L}$&
$\frac{2}{3}\sin^{2}\theta_{W}$&
&
$\frac{5-9\sin^{2}\theta_{W}}{6\sqrt{2-3\sin^{2}\theta_{W}}}$&
&
$-\cos\theta_{W}$\tabularnewline
&
&
&
&
&
\tabularnewline
$D_{1L}^{\prime},D_{2L}^{\prime}$&
$\frac{2}{3}\sin^{2}\theta_{W}$&
&
$\frac{5-9\sin^{2}\theta_{W}}{6\sqrt{2-3\sin^{2}\theta_{W}}}$&
&
$\cos\theta_{W}$\tabularnewline
&
&
&
&
&
\tabularnewline
$U_{3L}$&
$-\frac{4}{3}\sin^{2}\theta_{W}$&
&
$\frac{-1+9\sin^{2}\theta_{W}}{6\sqrt{2-3\sin^{2}\theta_{W}}}$&
&
$\cos\theta_{W}$\tabularnewline
&
&
&
&
&
\tabularnewline
$U_{3L}^{\prime}$&
$-\frac{4}{3}\sin^{2}\theta_{W}$&
&
$\frac{-1+9\sin^{2}\theta_{W}}{6\sqrt{2-3\sin^{2}\theta_{W}}}$&
&
$-\cos\theta_{W}$\tabularnewline
&
&
&
&
&
\tabularnewline
&
&
&
&
&
\tabularnewline
\hline 
&
&
&
&
&
\tabularnewline
\end{tabular}
\end{table}

\section{Seesaw Mechanism }

Now, let us inspect the gauge-invariant Ld of the Yukawa sector for
leptons. In our approach, it reads 

\begin{equation}
\mathcal{L}_{Y}^{lept}=G_{\alpha\beta}\bar{f}_{\alpha L}\left(\phi^{(4)}e_{\alpha L}^{c}+S_{R}f_{\beta L}^{c}+S_{D}f_{\beta L}^{c}+S_{D}^{\prime}f_{\beta L}^{c}\right)+H.c.\label{Eq. 21}\end{equation}
where $S$ matrices are defined as follows $S_{R}=\phi^{-1}(\phi^{(1)}\otimes\phi^{(2)}+\phi^{(2)}\otimes\phi^{(1)})\sim(\mathbf{1},\mathbf{10},1/2)$,
$S_{D}=\phi^{-1}(\phi^{(2)}\otimes\phi^{(3)}+\phi^{(3)}\otimes\phi^{(2)})\sim(\mathbf{1},\mathbf{10},1/2)$,
$S_{D}^{\prime}=\phi^{-1}(\phi^{(1)}\otimes\phi^{(3)}+\phi^{(3)}\otimes\phi^{(1)})\sim(\mathbf{1},\mathbf{10},1/2)$. 

After the SSB the first term in Eq. (\ref{Eq. 21}) supplies the masses
for all the charged leptons: $m(e)=A\left\langle \phi^{(4)}\right\rangle $,
$m(\mu)=B\left\langle \phi^{(4)}\right\rangle $, $m(\tau)=C\left\langle \phi^{(4)}\right\rangle $.
Obviously, $A=G_{11}$, $B=G_{22}$, $C=G_{33}$. 

The following 3 terms in Eq. (\ref{Eq. 21}) - when boosting to the
unitary gauge - will contribute to the mass of the neutrinos if the
first two positions in the lepton 4-plet gain a particular semnification.
A very strange - but meaningful outcome! - occurs for $N_{\alpha L}$
and $N_{\alpha L}^{\prime}$ when inspecting the Table containing
the fermion couplings to the neutral currents. As one expects, they
do not couple to the SM $Z$ boson, while their couplings to $Z^{\prime}$
are identical. Regarding their couplings to $Z^{\prime\prime}$ they
are identical up to a sign. This state of affairs entitles us to consider
that these neutral fermions could well be interpreted as 3 flavors
of right-handed neutrinos and their correspondig charge conjugates,
in the manner $N_{\alpha L}\equiv\nu_{\alpha R}$ and $N_{\alpha L}^{\prime}\equiv(\nu_{\alpha R})^{c}$.
With this identification one can easily observe that the Yukawa Ld
(\ref{Eq. 21}) leads (after the SSB) straightforwardly to the canonical
see-saw terms in the neutrino sector 

\begin{equation}
\mathcal{L}_{Y}^{\nu}=\mathcal{L}_{Y}^{D}(a)+\mathcal{L}_{Y}^{D^{\prime}}(a)+\mathcal{L}_{Y}^{R}(a),\label{Eq. 22}\end{equation}
which develop the following matrix

\begin{equation}
M_{\alpha\beta}^{M+D}=G_{\alpha\beta}\left(\begin{array}{ccc}
m_{R} &  & m_{D}^{T}\\
\\m_{D} &  & 0\end{array}\right)\label{Eq.23}\end{equation}
since the specific Dirac Yukawa Ld stands as $\mathcal{L}_{Y}^{D}=-m_{D}\bar{\psi^{c}}\psi+H.c.$
and the Majorana mass term as $\mathcal{L}_{Y}^{M}=-\frac{1}{2}m_{M}\bar{\psi^{c}}\psi+H.c.$ 

From this point on, our parametrization of the scalar sector plays
a crucial role in working out the see-saw mechanism. Assuming the
parameter outcome (\ref{Eq.9}) and the re-definition of the scalar
fields presented at the end of Sec. 2.3, one obtains:

\begin{equation}
M{}_{\alpha\beta}=G_{\alpha\beta}\left(\begin{array}{ccc}
2(1-a) &  & \sqrt{a(1-a)(1+\tan^{2}\theta_{W})}\\
\\\sqrt{a(1-a)(1+\tan^{2}\theta_{W})} &  & 0\end{array}\right)\left\langle \phi\right\rangle .\label{Eq. 24}\end{equation}

If the most suitable case requires the parameter $a\rightarrow0$,
the above see-saw mechanism exhibits the eigenvalue-matrix:

\begin{equation}
M(\nu_{L})=\frac{1}{2}a(1+\tan^{2}\theta_{W})\left(\begin{array}{ccc}
A & D & E\\
D & B & F\\
E & F & C\end{array}\right)\left\langle \phi\right\rangle ,\label{Eq. 25}\end{equation}
for the left handed-neutrinos, and 

\begin{equation}
M(\nu_{R})=2(1-a)\left(\begin{array}{ccc}
A & D & E\\
D & B & F\\
E & F & C\end{array}\right)\left\langle \phi\right\rangle ,\label{Eq.26}\end{equation}
for their right-handed partners. The Yukawa couplings in the above
expressions are $A=G_{ee}$, $B=G_{\mu\mu}$, $C=G_{\tau\tau}$, $D=G_{e\mu}$,
$E=G_{e\tau}$, $F=G_{\mu\tau}$ in our notation, and they should
disappear by solving an appropriate set of equations for different
mixing angles choices. 

The physical neutrino mass issue can be addressed if we consider first
neutrino mixing (for details see Refs. \cite{key-11} - \cite{key-13}
) The unitary mixing matrix $U$ (with $U^{+}U=1$) links the gauge-flavor
basis to the physical basis of massive neutrinos in the manner:

\begin{equation}
\nu_{\alpha L}(x)=\sum_{i=1}^{3}U_{\alpha i}\nu_{iL}(x)\label{Eq.27}\end{equation}
where $\alpha=e,\mu,\nu$ (corresponding to neutrino gauge eigenstates),
and $i=1,2,3$ (corresponding to massive physical neutrinos with masses
$m_{i}$). The mixing matrix $U$ that diagonalizes the mass matrix
$U^{T}MU=m_{ij}\delta_{j}$ has in the standard parametrization the
form: 

\begin{equation}
U=\left(c\begin{array}{ccc}
c_{2}c_{3} & s_{2}c_{3} & s_{3}e^{-i\delta}\\
-s_{2}c_{1}-c_{2}s_{1}s_{3}e^{i\delta} & c_{1}c_{2}-s_{2}s_{3}s_{1}e^{i\delta} & c_{3}s_{1}\\
s_{2}s_{1}-c_{2}c_{1}s_{3}e^{i\delta} & -s_{1}c_{2}-s_{2}s_{3}c_{1}e^{i\delta} & c_{3}c_{1}\end{array}\right)\label{Eq.28}\end{equation}
where the substitutions $\sin\theta_{23}=s_{1}$, $\sin\theta_{12}=s_{2}$,
$\sin\theta_{13}=s_{3}$, $\cos\theta_{23}=c_{1}$, $\cos\theta_{12}=c_{2}$,
$\cos\theta_{13}=c_{3}$ for the mixing angles have been made, and
$\delta$ is the CP phase. Bearing in mind that $TrM(\nu_{L})=\sum_{i}m_{i}$
and phenomenological values $m_{i}$ of neutrino masses are severely
limited to few $eV,$one obtains: $\sum_{i}m_{i}=\frac{1}{2}a(1+\tan^{2}\theta_{W})\left\langle \phi\right\rangle (A+B+C)$.
That is

\begin{equation}
TrM(\nu_{L})=\frac{1}{\sqrt{2}}\left(\frac{1+\tan^{2}\theta_{W}}{\sqrt{1-\tan^{2}\theta_{W}}}\right)m(\tau)\left[1+\frac{m(\mu)}{m(\tau)}+\frac{m(e)}{m(\tau)}\right]\sqrt{a}\label{Eq.29}\end{equation}
 With its good approximation:

\begin{equation}
\sum_{i}m_{i}\simeq\frac{1}{\sqrt{2}}\left(\frac{1+\tan^{2}\theta_{W}}{\sqrt{1-\tan^{2}\theta_{W}}}\right)m(\tau)\sqrt{a}\label{Eq.30}\end{equation}
where we neglected the small ratios $m(\mu)/m(\tau)\sim0.05$ and
$m(e)/m(\tau)\sim0.0002$ in Eq.(\ref{Eq.29}) and exploited $m(\tau)=C\frac{\sqrt{a(1-\tan^{2}\theta_{W})}}{\sqrt{2}}\left\langle \phi\right\rangle $.
With this result (taking into account the PDG results \cite{key-68})
one can estimate the range of the free parameter $a$ in order to
match the observed tiny masses ($\sim1eV$) in the left-handed neutrino
spectrum. It has to be $a\sim0.25\times10^{-18}$ corresponding to
a mass scale $m\sim1.6\times10^{11}$GeV. (The latter was inferred
from Eq.(\ref{Eq10}) in order to ensure ${m}(W)=80.4$GeV. Under
these circumstances, right-handed neutrinos must exhibit masses in
the range $\sim1.2\times10^{10}$GeV, unaccesible yet to a direct
observation. 

Regarding the mixing angles, the neutrino mass hierarchy (normal,
inverted or degenerate) and its splitting, or possible additional
symmetries (such as $L_{e}-L_{\mu}-L_{\tau}$) for a neutrino mass
matrix including diagonal entries that are proportional to the charged
lepton masses was treated in Ref. \cite{key-69}. Those results are
suitable for the model of interest in this paper, since our diagonal
entries here exhibit the same proportionality. 

We mention also that some new bosons ($Y$, $Z'$ and $Z''$) gain
masses at the $10^{11}$GeV level. But this is not a contradiction,
since our point of departure in our analysis consisted in decoupling
of the heavier neutral boson ($Z''$) from its two companions ($Z$
and $Z'$).

\section{Conclusions}

In conclusion, in this paper we have worked out the neutrino mass
issue in a 3-4-1 electro-weak model without exotic electric charges,
proving that the canonical see-saw mechanism can naturally arise -
without resorting to any supplemental ingredients! - by just exploiting
the general method of treating gauge models with high symmetries.
This assumes a geometrical approach - given by a proper parameter
set in the Higgs sector - combined with the redefinition of the scalar
multiplets and a particular gauge fixing (we work in unitary gauge
from the very beginning for some nedeed scalar 10-plets - \emph{i.e.}
matrices $S$ in Eq.(\ref{Eq. 21}, constructed as tensor products
out of the existing 4-plets). This procedure leads straightforwardly
to the one-parameter ($a$) see-saw mechanism giving the right order
of magnitude for the left-handed neutrinos $\sim$eV when the mass
scale of the whole model lies in the range $m\sim10^{11}$GeV. The
SM phenomenology is not disturbed by this mathematical approach, since
all the masses and couplings of the SM particles - namely, leptons
leptons $e$, $\mu$, $\tau$, quarks $u$, $d$,$s$,$c$,$t$,$b$,
and bosons $W$, $Z$ plus the massless $A_{em}$- computed through
our method come out at their established values.

\end{document}